\documentclass[conference]{IEEEtran}
\IEEEoverridecommandlockouts
\usepackage{cite}
\usepackage{amsmath,amssymb,amsfonts}
\usepackage{algorithmic}
\usepackage{graphicx}
\usepackage[labelfont=bf,textfont=bf]{caption}
\usepackage{textcomp}
\usepackage{xcolor}
\usepackage[hidelinks]{hyperref}  % No colored boxes
\usepackage{multirow} 
\usepackage{enumitem}

\def\BibTeX{{\rm B\kern-.05em{\sc i\kern-.025em b}\kern-.08em
    T\kern-.1667em\lower.7ex\hbox{E}\kern-.125emX}}

\author{
    \IEEEauthorblockN{
        Akash Poptani\IEEEauthorrefmark{2} \hspace{2em}
        Alireza Khadem\IEEEauthorrefmark{2} \hspace{2em}
        Scott Mahlke\IEEEauthorrefmark{2}
    }
    \IEEEauthorblockA{
        akashpt@umich.edu \hspace{2em}
        arkhadem@umich.edu \hspace{2em}
        mahlke@umich.edu
    }
    \\
    \IEEEauthorblockN{
        Jonah Miller\IEEEauthorrefmark{4} \hspace{2em}
        Joshua Dolence\IEEEauthorrefmark{4} \hspace{2em}
        Galen Shipman\IEEEauthorrefmark{4} \hspace{2em}
        Reetuparna Das\IEEEauthorrefmark{2}
    }
    \IEEEauthorblockA{
        jonahm@lanl.gov \hspace{2em}
        jdolence@lanl.gov \hspace{2em}
        gshipman@lanl.gov \hspace{2em}
        reetudas@umich.edu
    }
    \\
    \IEEEauthorblockA{
        \IEEEauthorrefmark{2}University of Michigan \hspace{3em}
        \IEEEauthorrefmark{4}Los Alamos National Laboratory
    }
}

\begin{document}

% \title{Characterizing Adaptive Mesh Refinement on Heterogeneous Platforms with Parthenon-VIBE\\

\title{Characterizing Adaptive Mesh Refinement on Heterogeneous Platforms with Parthenon-VIBE\\
% \thanks{ This material is based upon work supported by the Department of Energy, National Nuclear Security Administration under contract number 89233218CNA000001. 
%  LA-UR-25-ABCDE Approved for public release; distribution is unlimited.}
}

% \author[1]{Akash Poptani}
% \author[1]{Alireza Khadem}
% \author[1]{Scott Mahlke}
% \author[2]{\\Jonah Miller}
% \author[2]{Joshua Dolence}
% \author[2]{Galen Shipman}
% \author[1]{Reetuparna Das}

% \affil[1]{University of Michigan \\ 
% \texttt{\{akashpt, arkhadem, mahlke, reetudas\}@umich.edu}}
% \affil[2]{Los Alamos National Laboratory \\ 
% \texttt{\{jonahm, jdolence, gshipman\}@lanl.gov}}

\maketitle

\begin{abstract}
Hero-class HPC simulations rely on Adaptive Mesh Refinement (AMR) to reduce compute and memory demands while maintaining accuracy. This work analyzes the performance of Parthenon, a block-structured AMR benchmark, on CPU-GPU systems. We show that smaller mesh blocks and deeper AMR levels degrade GPU performance due to increased communication, serial overheads, and inefficient GPU utilization. Through detailed profiling, we identify inefficiencies, low occupancy, and memory access bottlenecks. We further analyze rank scalability and memory constraints, and propose optimizations to improve GPU throughput and reduce memory footprint. Our insights can inform future AMR deployments on Department of Energy's upcoming heterogeneous supercomputers.
\end{abstract}

\vspace{-14pt}
\section{Introduction}

% AMR Definition and why is it important
Hero-class High-Performance Computing (HPC) simulations require over half a petabyte of memory capacity and take weeks to months on state-of-the-art supercomputers~\cite{9984818}. Adaptive Mesh Refinement (AMR) is a widely adopted and efficient computational technique for these simulations. Instead of uniformly simulating the entire computational grid at the highest resolution, which demands substantial computational and memory resources, AMR dynamically refines the mesh in regions requiring greater detail, while preserving a coarser grid in less critical areas. This approach enables high simulation accuracy while significantly reducing the overall compute and memory requirements~\cite{amrclassic}. AMR has proven effective in a range of scientific domains, including astrophysics \cite{castro, Gamer, Enzo}, fluid dynamics \cite{cholla, athena++, kathena, Flash, RAMSES, warpx}, weather and climate modeling \cite{climate1, climate2, climate3, climate4}, and combustion flows \cite{PeleC, rendleman1998amr, sitaraman2021adaptive}.
 % and reactive 

% GPU is important, what we do in the paper
Parthenon is a \textit{block-structured AMR} framework~\cite{chombo, AMReX, parthenon} designed for solving Partial Differential Equations (PDEs) in multi-physics simulations.
As the HPC ecosystem transitions to heterogeneous CPU-GPU architectures, Parthenon adopts the Kokkos abstraction layer~\cite{kokkos1, kokkos2} to enable flexible integration with the CPU and GPU backends. Although GPUs provide high throughput, sustaining efficiency requires careful tuning of mesh resolution and refinement parameters, which vary widely between simulations. This work characterizes the CPU and GPU performance of \textit{Parthenon-VIBE}, a benchmark that solves the Vector Inviscid Burgers' Equation\cite{vibe}. Parthenon-VIBE serves as a key proxy application for ATS-5, the Department of Energy’s next-generation supercomputer~\cite{ats5github, ats5timeline, ats5guide}, scheduled for deployment in 2027. 

\begin{figure}[t]
    \centering
    \includegraphics[width=\linewidth]{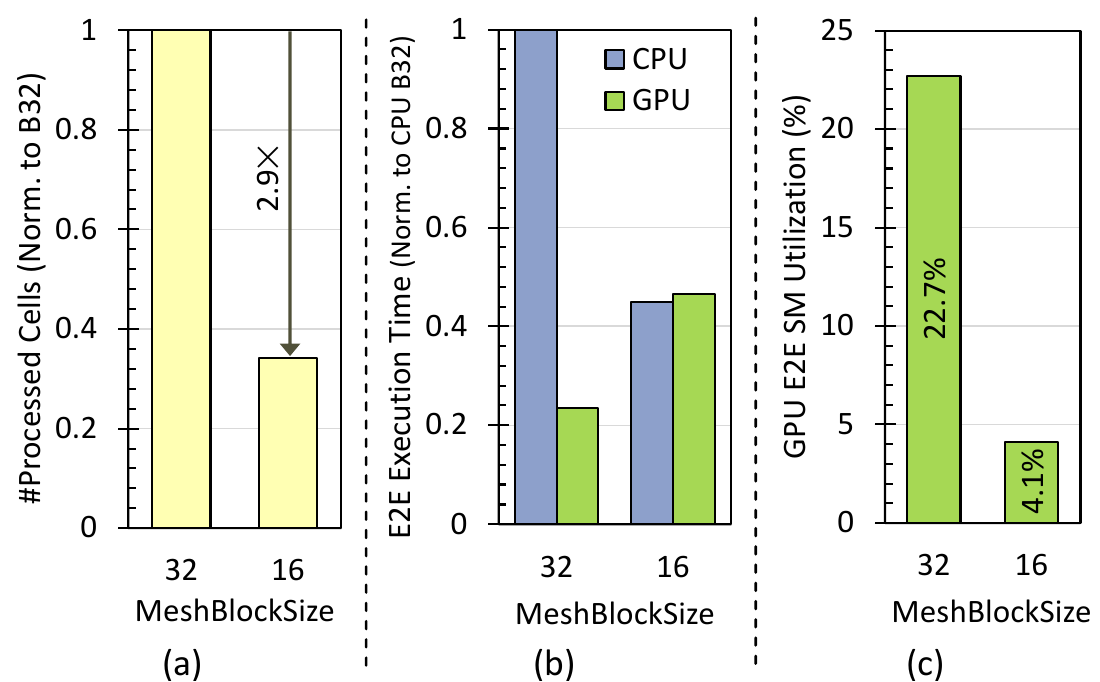}
    \caption{Effect of mesh block size on Parthenon performance:  (a) Smaller mesh blocks reduce computation through finer control on resolution. (b) H100 GPU performance degrades with smaller blocks, matching or lagging behind a 96-core Sapphire Rapids CPU. (c) H100 utilization drops significantly with smaller mesh blocks.}
    \label{fig:motivation}
\end{figure}

% Block-structured AMR partitions the mesh grid into sub-blocks and refines them independently across multiple AMR levels. Mesh blocks can be independently refined for higher resolution in the next AMR level. Finer blocks reduce the compute and memory requirements by adjusting the resolution where it is truly needed. For example, Fig.~\ref{fig:motivation}(a) shows that using a mesh block size of 16 reduces the number of processed cells by 2.9$\times$ compared to a block size of 32. Motivated by such efficiency, frameworks like RAGE~\cite{rage} adopt cell-based AMR, with the smallest possible block size equal to one cell. However, finer mesh blocks and deeper AMR levels introduce performance challenges due to irregular memory access, synchronization overhead, and high communication cost. As shown in Fig.~\ref{fig:motivation}(b), an H100 GPU can perform similarly or worse than a 96-core Sapphire Rapids CPU when using smaller mesh blocks and more AMR levels. Moreover, Fig.~\ref{fig:motivation}(c) demonstrates that GPU throughput utilization drops sharply under these conditions, leading to the waste of costly GPU resources.

Block-structured AMR partitions the mesh grid into sub-blocks, known as \textit{mesh blocks}, which can be independently refined and divided into multiple smaller mesh blocks in the next AMR level. Using a smaller mesh block size reduces the compute and memory requirements by adjusting the resolution where it is truly needed. For example, Fig.~\ref{fig:motivation}(a) shows that using a mesh block size of 16 reduces the number of processed cells by 2.9$\times$ compared to a block size of 32. Motivated by such efficiency, frameworks like RAGE~\cite{rage} adopt cell-based AMR, with the smallest possible block size equal to one cell. 

However, \textit{finer blocks and deeper AMR levels} significantly increase the number of mesh blocks, resulting in more serial overhead for block management, higher communication and synchronization costs, and more irregular memory access patterns.
Additionally, these higher numbers of mesh blocks are processed with low SM utilization on GPU. 
%\textcolor{red}{
% Additionally, smaller mesh blocks lead to an increased number of kernel launches, which can raise kernel execution overhead.
 % }
% \textcolor{red}{Additionally, smaller mesh blocks require more iterations within Kokkos kernels, which increases kernel execution time.} 
Under these conditions, Fig.~\ref{fig:motivation}(b) shows that an H100 GPU can perform similarly or worse than a 96-core Sapphire Rapids CPU. Moreover, Fig.~\ref{fig:motivation}(c) demonstrates that GPU utilization drops sharply, leading to the waste of costly GPU resources.

% We find that at fine resolutions, as Mesh block sizes decrease and AMR levels increase, GPU performance declines due to several factors: the higher number of mesh blocks results in higher communication overhead; a larger portion of runtime is spent on serial tasks like block management; and \textcolor{red}{smaller mesh blocks require more iterations within Kokkos kernels, which increases kernel execution time.} Together, these effects reduce overall end-to-end  GPU utilization and throughput in fine-grain AMR scenarios. 

Our timing analysis identifies these key performance bottlenecks and illustrates the runtime distribution across major code components. We show that GPU performance degradation can be alleviated by increasing the number of MPI ranks per GPU, which helps overcome Amdahl's bottlenecks in serial regions of the code. However, scalability is limited by the GPU's memory capacity. As the number of ranks increases, the memory consumed by MPI communication buffers and the Open MPI driver grows significantly, leading to out-of-memory (OOM) errors.

% Our timing analysis highlights these major bottlenecks and the distribution of runtime between the main components. We show that GPU performance degradation can be mitigated by increasing the number of MPI ranks per GPU, overcoming Amdahl's bottlenecks in serial portions of the code. However, rank scalability is constrained by the limited memory capacity of the GPU device. When increasing the number of ranks, MPI communication buffers and driver consume substantial memory capacity, and result in out of memory errors.

We conduct a detailed microarchitectural analysis covering GPU throughput, occupancy, warp utilization, memory bandwidth utilization, and arithmetic intensity. Our findings reveal low SM occupancy due to high register requirements. We discuss unoptimized Kokkos CUDA block configurations, which lead to redundant computations and control flow divergence. Although the kernels are memory-bound due to low arithmetic intensity, memory bandwidth utilization remains low because of the sparse memory accesses. On the CPU side, opcode analysis shows a high proportion of vector instructions in kernels and load and store instructions in the serial portion of the code.

Finally, we present a set of hardware-aware software recommendations designed to enhance GPU scalability and minimize the application's memory footprint. These include optimizing memory allocation, increasing GPU occupancy, and exploring support for finer-grained offloading of serial components.

%Memory footprint can be reduced through kernel fusion, which combines consecutive GPU kernels that produce and consume intermediate data. By fusing these dependent kernels into a single computational kernel, the need to allocate and store temporary intermediate variables in memory is eliminated.
%\todo{GPU is helped more - decide if we want to put the figure or not- batching tradeoff - use it as an example - dont put the figure}
% \begin{figure}
%     \centering
%     \includegraphics[width=1\linewidth]{buffer-block-packing-overhead.png}
%     \caption{Enter Caption}
%     \label{fig:enter-label}
% \end{figure}

%\ignore{

 In summary, the key contributions of this work are:

\begin{itemize}[topsep=0pt, itemsep=0ex]
    \item A detailed performance characterization of the Parthenon AMR framework on CPUs and GPUs, across varying mesh sizes, mesh block sizes, and AMR levels, using Burgers benchmark.

    \item Identification of the performance degradation patterns on GPUs, including the impact of small mesh blocks and deep AMR hierarchies on kernel efficiency and overall GPU utilization.
    
    \item Analysis of serial bottlenecks that limit GPU performance, supported by timing breakdowns, microarchitectural statistics, and hotspot profiling across CPU and GPU executions.
    
    \item Evaluation of rank scaling behavior per GPU, including memory constraints, and discussion of rank-performance trade-offs.
    
    \item We suggest hardware-aware software recommendations to improve AMR performance on GPUs, including rank configuration, memory optimizations, and offload strategies for serial regions.
    \end{itemize}
    
% \vspace{-13pt}
\section{Background}
% \vspace{-8pt}

\subsection{Adaptive Mesh Refinement (AMR)}

AMR is a computational technique that dynamically adjusts mesh resolution in regions where the simulation domain exhibits significant variation, enabling higher accuracy where needed while minimizing computational and memory capacity costs. The mesh is a logical representation of the discretized physical domain. For example, in simulating heat conduction in a rod heated at one end, the region of interest near the heat source exhibits a steep temperature gradient, while other regions show more gradual changes. AMR applies a fine mesh within this region of interest to capture sharp variations accurately and a coarse mesh farther away to reduce computation and memory. This approach improves accuracy compared to a uniformly coarse mesh and reduces computational and memory expenses relative to a uniformly fine mesh. AMR inherently supports hierarchical parallelism by organizing computations into multiple refinement levels. Each level corresponds to a distinct resolution and can be processed concurrently, enhancing scalability and resource utilization.
\vspace{-3pt}
\subsection{Tree Based AMR}
\vspace{-1pt}

Tree-based AMR uses a hierarchical tree structure to represent the mesh refinement levels in the computational domain. Each node in the tree corresponds to a MeshBlock, and the tree explicitly defines the parent-child relationships created during refinement or derefinement. The entire tree is rebuilt every time refinement or derefinement occurs to maintain consistency. One key advantage of this approach is that any spatial location in the domain is covered by exactly one MeshBlock, ensuring no overlap between parent and child blocks in physical space. Consequently, neighbor relationships exist only between MeshBlocks at the same refinement level, with no spatial parent-child ones. This structure is typically represented as a binary tree in one dimension, a quadtree in two dimensions, or an octree in three dimensions, where each parent node subdivides into two, four, or eight child nodes, respectively.

\begin{figure}[h]
    \centering
    \includegraphics[width=\linewidth]{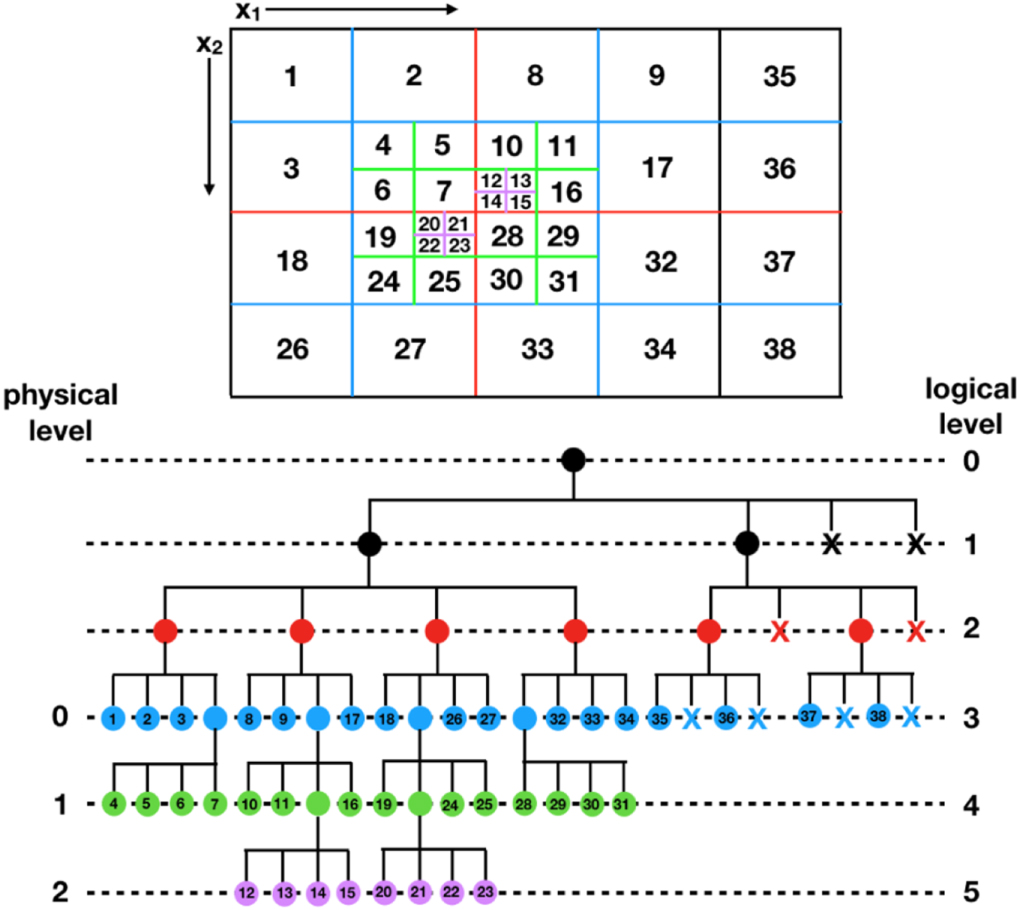}
    \caption{Illustration of a quadtree structure used in 2D tree-based AMR. Different colors indicate refinement levels, with empty leaves (marked by X) representing regions outside the physical domain~\cite{parthenon}.}
    \label{fig:quadtree}
\end{figure}

The quadtree, illustrated in Fig.\ref{fig:quadtree} represents the hierarchical refinement structure. The root node at logical level 0 corresponds to the entire computational domain covered by the tree, which may include regions beyond the physical domain due to the symmetrical subdivision inherent in the quadtree. The root node at physical level 0 corresponds exactly to the physical domain of interest. At the top, the physical grid is shown with MeshBlocks arranged uniformly in a 5 by 4 layout. This uniform grid corresponds to logical level 3 in the quadtree, which matches the base physical level 0. Empty leaves (marked by X) represent regions outside the physical domain. Refinement introduces additional logical levels, corresponding to finer MeshBlocks nested within coarser ones. These refinements are shown up to two levels deep, with child nodes representing refined MeshBlocks. Node colors differentiate refinement levels and illustrate the hierarchical structure. The quad tree enables efficient management of mesh blocks and supports hierarchical refinement while preserving a clean spatial partitioning of the domain.

\subsection{Parthenon}
Parthenon is a performance-portable, block-structured AMR framework supporting both CPU and GPU architectures. Building on Athena++~\cite{athena++} (CPU-only) and K-Athena~\cite{kathena} (CPU/GPU via Kokkos), Parthenon extends from a domain-specific astrophysical code to a generalized, extensible AMR framework for multiple scientific fields. Its GPU support also relies on the \textbf{Kokkos} abstraction layer~\cite{kokkos1, kokkos2}. Kokkos is an open-source, C++ template-based programming model that facilitates efficient execution on CPUs, GPUs, and other accelerators. Analysis of Kokkos kernels provides us insights into the portion of code offloaded to GPU and data parallel portions on CPU. %\todo{there is no OMP parallel portion on CPU. at least we don't analyze it. maybe vectorized?} .
% data parallel is broader, since we have 1 thread per rank, there is no OMP in our experiments, TBD in future as Galen suggested.
We define the \textbf{serial portion} as code that lies outside these Kokkos kernels.

Parthenon optimizes data management on GPU architectures by allocating all simulation data directly in GPU device memory, thereby minimizing costly data transfers between host and device. It supports logical packing of variables and mesh blocks, reducing kernel launch overhead~\cite{parthenon}. Communication overhead in multi-node simulations is minimized through asynchronous, one-sided MPI calls. Parthenon employs a hierarchical task-based execution model, enabling fine-grained parallelism with controlled task granularity. To maintain consistency and optimize communication, restriction averages data from finer to coarser mesh levels. Parthenon applies restriction before sending data from finer to coarser blocks during MPI communication, reducing data volume, and improving performance.

%\subsection{Parthenon Structure}

\begin{figure}[t]
    \centering
    \includegraphics[width=1\linewidth]{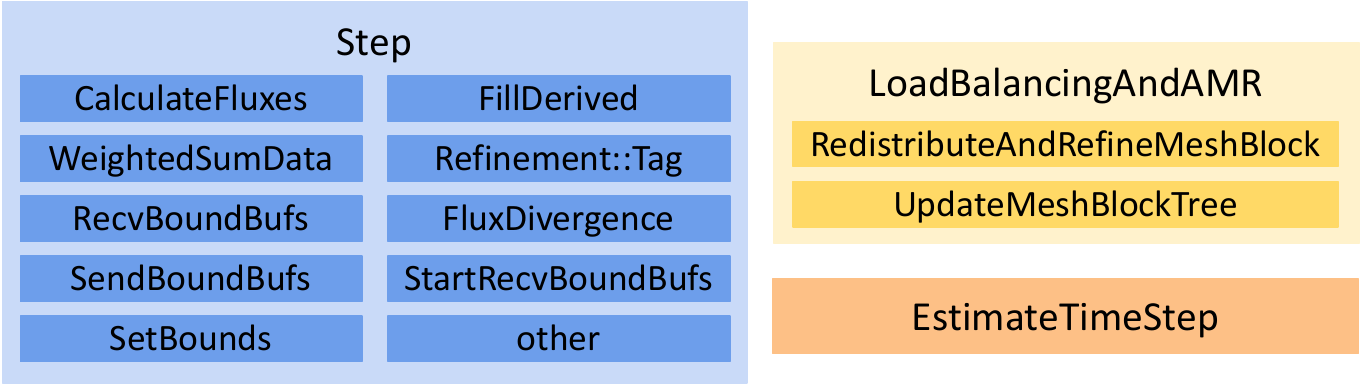}
    \caption{Main (sub)functions of Parthenon timestep loop.}
    \label{fig:topdown}
\end{figure}

Most scientific simulations evolve over time by repeatedly executing a fixed sequence of computations, known as timestepping, until reaching a certain end time. In Parthenon, this loop iteratively advances the simulation by invoking three key functions: \texttt{Step}, \texttt{LoadBalancingAndAMR}, and \texttt{EstimateTimeStep}, which are depicted in Fig.~\ref{fig:topdown}. Consider a stone dropped into still water, creating ripples that spread outward over time. The \texttt{Step} function updates the water surface height and velocity as the ripples evolve. \texttt{LoadBalancingAndAMR} refines the mesh near steep wavefronts and coarsens it where the surface is calm. \texttt{EstimateTimeStep} adjusts the timestep size, using smaller steps when waves are sharp and larger ones as they smooth, maintaining numerical stability.

The \texttt{CalculateFluxes} function is responsible for the core numerical computations in the time integration of the PDEs. It first performs a fifth-order Weighted Essentially Non-Oscillatory (WENO5)\cite{weno5} reconstruction using stencil computations to approximate the physical variables at cell faces. Using these reconstructed face values, the function then applies a Riemann solver to compute the fluxes across cell interfaces. These fluxes are essential inputs for updating conserved quantities in the simulation.

Parthenon performs \texttt{FluxCorrection} at fine-coarse mesh boundaries to prevent conservation errors, such as artificial gains or losses of mass, momentum, or energy, that arise because the aggregate of fluxes computed on finer mesh cells does not exactly match the flux computed on the corresponding coarse cell. This operation is carried out through an interblock communication step, which leverages the same mechanism as ghost zone exchanges but applies only to flux fields. Subsequently, \texttt{FluxDivergence} calculates the divergence of fluxes of conserved variables.

For time integration, Parthenon uses the Runge-Kutta method~\cite{rk2}. The first stage, \texttt{AverageIndependentData}, computes an initial estimate of the solution advancement based on the current state. The second stage, \texttt{UpdateIndependentData}, refines this estimate by incorporating information from the updated state later in the timestep. For example, to track how far a particle moves over a short interval, the first stage estimates the distance using the particle’s initial speed, while the update stage improves this estimate by averaging the initial and final speeds. This approach enhances accuracy and ensures numerical stability during time evolution.

\subsection{Ghost Cell Communication}
Ghost cell communication is critical in Parthenon to support stencil computations like those performed in \texttt{CalculateFluxes}. Since stencil operations require data from neighboring cells—including those on adjacent mesh blocks—ghost cells at block boundaries must be updated every timestep to ensure consistent flux calculations.

The communication process involves four major functions. First, \texttt{StartReceiveBoundBufs} prepares buffers to receive incoming ghost cell data asynchronously. Following this, \texttt{SendBoundBufs} prepares and sends data to neighboring mesh blocks. It handles metadata and buffer caches to ensure data consistency, performs restriction of fine mesh data to coarser mesh regions (GPU offloaded), efficiently packs variable data into communication buffers (GPU offloaded where applicable; see Section 3.6 in the Parthenon paper \cite{parthenon}), and initiates non-blocking sends or local memory copies to transmit the data.

Once data is in transit, \texttt{ReceiveBoundBufs} monitors and manages the arrival of incoming data. For local communication within the same MPI rank, data is marked as received immediately. For remote communication, it uses \texttt{MPI\_Iprobe} to non-blockingly check for incoming messages and gently nudge the MPI library to progress communication. When messages are detected, \texttt{MPI\_Test} confirms the completion of specific receive operations. After confirming data arrival, memory is allocated as necessary, and the data is prepared for subsequent computation.

Finally, \texttt{SetBounds} completes the communication cycle by transferring the received ghost cell data from buffers into the simulation’s primary variables. It updates buffer metadata, launches GPU-accelerated kernels to unpack and copy data, marks communication buffers as stale to prepare for future exchanges. GPU fences ensure completion of all memory operations before proceeding.

\subsection{\texttt{LoadBalancingAndAMR}}

The \texttt{LoadBalancingAndAMR} function performs two major tasks: \textit{First}, \texttt{RedistributeAndRefineMeshBlocks} dynamically reorganizes mesh blocks to maintain load balance across MPI ranks and accommodate changes in mesh refinement. It updates the mesh block list to include newly refined or derefined blocks, computes workload costs—based on estimated computational expense per block—to guide load balancing.
%, and redistributes blocks across MPI ranks to evenly distribute the computational workload. 
When blocks are refined, derefined, or assigned to different ranks, the corresponding data must be communicated. This data transfer utilizes Parthenon’s ghost cell communication mechanisms. Additionally, the function performs prolongation and restriction operations to synchronize solution variables between coarse and fine mesh blocks. It also rebuilds neighbor relationships and boundary buffers through calls to \texttt{BuildTagMapAndBoundaryBuffers} and \texttt{SetMeshBlockNeighbors} to maintain mesh connectivity and communication patterns. \textit{Second}, \texttt{UpdateMeshBlockTree} collects refinement and derefinement flags from all mesh blocks, aggregates these requests across MPI ranks via an “All Gather” operation, and determines which blocks require refinement or coarsening. It then updates the mesh block hierarchy by manipulating the binary, quad, or octree structure organizing the mesh blocks.

\subsection{\texttt{MeshBlockSize}, \texttt{\#AMR Levels} and Rules}

A Mesh is composed of several MeshBlocks. A MeshBlock is defined as a regular array of cells representing a subvolume of the computational domain. It also serves as the fundamental granularity for refinement. The size of each MeshBlock (\texttt{MeshBlockSize}) determines the resolution at which refinement occurs. The parameter \texttt{\#AMR Levels} sets the maximum depth to which the refinement tree can grow.

Refinement flexibility is enabled by the choice of \texttt{MeshBlockSize} and the maximum \texttt{\#AMR Levels}. Smaller MeshBlocks enable more precise spatial refinement, focusing computational effort on specific regions and resulting in finer blocks distributed sparsely across the domain. Likewise, a higher \texttt{\#AMR Levels} allows for deeper, more localized refinement. Together, these factors facilitate \textit{sparse computation} by increasing resolution only in selected regions.

To ensure numerical stability and consistency, the $2:1$ refinement rule is enforced: neighboring MeshBlocks cannot differ by more than one refinement level. Additionally, the total mesh size in each spatial dimension must be an exact multiple of the corresponding MeshBlock size. This ensures that the entire mesh can be divided evenly into MeshBlocks, enabling efficient and organized AMR operations.

\subsection{Burgers Benchmark}
The Burgers benchmark solves a simplified partial differential equation (PDE) that retains key features characteristic of AMR workloads. It provides a representative performance problem for evaluating AMR frameworks. The benchmark solves the following equation for the velocity field \(\mathbf{u}\):
\[
\frac{\partial \mathbf{u}}{\partial t} + \nabla \cdot \left( \frac{1}{2} \mathbf{u} \mathbf{u} \right) = 0,
\]
while simultaneously evolving one or more passive scalar quantities \( q^i \) according to:
\[
\frac{\partial q^i}{\partial t} + \nabla \cdot (q^i \mathbf{u}) = 0.
\]
Additionally, an auxiliary quantity \( d \), resembling kinetic energy, is computed as:
\[
d = \frac{1}{2} q^0 \mathbf{u} \cdot \mathbf{u}.
\]
Parthenon-VIBE employs a Godunov-type finite volume scheme\cite{finitevolumescheme} with options for slope-limited linear or WENO5 reconstruction\cite{weno5}, HLL fluxes, and second-order Runge-Kutta time integration\cite{rk2}. The 3D solution of the Vector Inviscid Burgers’ Equation serves as a proxy application for block-structured AMR stencil operations, as demonstrated in LANL’s ATS 5 documentation \cite{ats5timeline}. This makes it an ideal choice for benchmarking AMR frameworks like Parthenon. We use WENO5 \cite{weno5} reconstruction for the analysis in this work. We set the refinement to occur every cycle, while derefinement operations are constrained by a minimum gap of 10 cycles between successive derefinements. Load balancing is performed every cycle to maintain computational efficiency.

%\vspace{-4mm}
\section{Characterization Methodology} \label{sec:param}

In this work, we focus on performance characterization of a \textbf{single-node}, as many of the inefficiencies in AMR on GPU/CPU architectures can be isolated to a single node. Multi-node inefficiencies, such as inter-node communication and load balancing, which can be exacerbated by GPU architectures, are left to future work.

%Although scaling to multiple nodes may still be necessary to achieve desired simulation turnaround times, improving performance on a single node reduces the total number of nodes required to meet the same performance goals \cite{lammpsiiswc}. We focus on single node as there are significant performance challenges on a single node for many AMR configurations. Future work can be multimode. We believe that characterizing single-node (CPU or GPU) performance is crucial for AMR-based benchmarks.

%This is orthogonal to multi-node studies.
%\vspace{-4mm}
%\subsection{Hardware}

The experiments were conducted on a single compute node equipped with both CPUs and GPUs. Each node contains 96 Intel Sapphire Rapids (SPR) CPU cores \cite{intel-sapphire-rapids} and up to 8 NVIDIA H100 GPUs \cite{nvidiah100}, as detailed in Tables~\ref{tab:cpu_specs} and~\ref{tab:gpu_specs}. For our experiments, we utilized the full 96-core Sapphire Rapids CPU and configurations with 1, 4, and 8 NVIDIA H100 GPUs to study scaling and performance.

\begin{table}[h!]
\centering
\caption{CPU Specifications}
\begin{tabular}{l c}
\hline
\textbf{Specification} & \textbf{Details} \\
\hline
Processor & Intel Xeon Platinum 8468 \\
 & (Sapphire Rapids) \\
Number of Cores & 96 \\
Number of Sockets & 2 \\
Base Frequency & 3.1 GHz \\
L1 Cache & 48 KB (L1d) + 32 KB (L1i) per core \\
L2 Cache & 2 MB (2048 KB) per core \\
L3 Cache & 105 MB (107520 KB) shared \\
Technology Node & 10 nm (Intel 7) \\
% Thermal Design Power (TDP) & 350W \\
Memory & 1.0 TiB DDR5 \\
Memory Bandwidth & 614.4 GB/s \\
Operating System & Red Hat Enterprise Linux 8.10 (Ootpa) \\
Kernel Version & Linux 4.18.0-553.50.1.el8\_10.x86\_64 \\
\hline
\end{tabular}
\label{tab:cpu_specs}
\end{table}
\begin{table}[h!]
\centering
\caption{GPU Specifications}
\begin{tabular}{l c}
\hline
\textbf{Specification} & \textbf{Details} \\
\hline
GPU Model & NVIDIA H100 \\
Streaming Multiprocessors (SMs) & 132 \\
Base Frequency & 1.98 GHz \\
Global Memory & 81,559 MiB HBM3 \\
Memory Bandwidth & 3.35TB/s \\ 
L1 Cache + Scratchpad & 256 KB \\
L2 Cache & 50 MB \\
Technology Node & TSMC 4N process node \\
% Thermal Design Power (TDP) & \todo{350W to 700W
% (configurable)} \\
\hline
\end{tabular}
\label{tab:gpu_specs}
\end{table}

%\subsection{Software}
The experiments use Parthenon version \textit{regression-gold-v25-86-g9b045aec7}, built with Kokkos and MPI to support both CPU and GPU execution.
For CPU builds, compiler optimizations targeted the Intel Sapphire Rapids architecture. The runtime environment modules were loaded accordingly: the CPU runs used the NVIDIA HPC compiler with Open MPI, while GPU runs utilized GCC 10.3.0 with Open MPI 4.1.6 configured for CUDA support.
This software setup ensures optimized performance and portability across the CPU and GPU hardware used in the experiments.

We use Kokkos Tools~\cite{kokkos_tools} to analyze the execution time and kernel memory usage distribution across code components. To study memory consumption outside Kokkos kernels, we collect memory allocation and deallocation traces along with call stacks using the NVIDIA Nsight Systems profiler. GPU microarchitectural characteristics are examined using the NVIDIA Nsight Compute profiler, while CPU instruction opcode distribution is analyzed with Intel PIN and MICA tools~\cite{MICA}.

\subsection{Figure of Merit (FOM)}
\noindent The primary metric used to evaluate Parthenon’s performance is \emph{zone-cycles per second}.
Zone-cycles is defined as the total number of mesh blocks processed in all simulation cycles multiplied by the block size in the \(x\), \(y\), and \(z\) dimensions:

\[
\text{zone-cycles} = N_\text{blocks} \times B_x \times B_y \times B_z
\]

% \todo{where \(N_\text{blocks}\) is the total number of mesh blocks processed, and \(B_x, B_y, B_z\) are the block sizes. 
% }
\[
\text{zone-cycles/sec} = \frac{\text{zone-cycles}}{\text{wall clock time (sec)}}
\]

This metric, equivalent to cell updates per second, directly measures computational throughput in AMR simulations. The objective is to maximize zone-cycles per second to indicate higher performance.

\section{Performance Characterization} \label{sec:performance}

% Insight 1: smaller mesh block and more AMR levels hurt the GPU performance and significantly increase memory capacity requirements.

% Why? smaller mesh block, more blocks, more serial part, less GPU part. Show with \% of offloaded GPU time.

% chart1: x axis mesh block size, y axis is zone/cycle, 4 lines 96 CPU and 1, 4, 8 GPUs

% chart2: x axis mesh block size, y axis is memory usage, 2 lines 96 CPU and 1 GPU

% chart3: x axis AMR levels, y axis is zone/cycle, 4 lines 96 CPU and 1, 4, 8 GPUs

% chart4: x axis AMR levels, y axis is memory usage, 2 lines 96 CPU and 1 GPU

\subsection{Mesh Size}

We performed a \textit{static scaling} experiment by fixing the hardware setup and increasing the problem size. \textbf{Mesh size was varied among \texttt{64}, \texttt{96}, \texttt{128}, \texttt{160}, \texttt{192}, and \texttt{256}} cells per dimension, which are shown on the X-axis of Fig.~\ref{fig:perf_mesh_size}. The \textbf{parameters \texttt{MeshBlockSize} and the \texttt{\#AMR levels} are kept constant at \texttt{16} and \texttt{3}}, respectively. The Y-axis shows performance measured by the figure of merit (FOM), given in zone-cycles per second. For each AMR configuration, we use different number of GPUs and MPI Ranks (denoted by \textbf{R}). The \textbf{BestR} configuration picks the most optimal number of MPI ranks for the GPU. This figure presents the following two takeaways:

\begin{figure}[t]
    \centering
    \includegraphics[width=\linewidth]{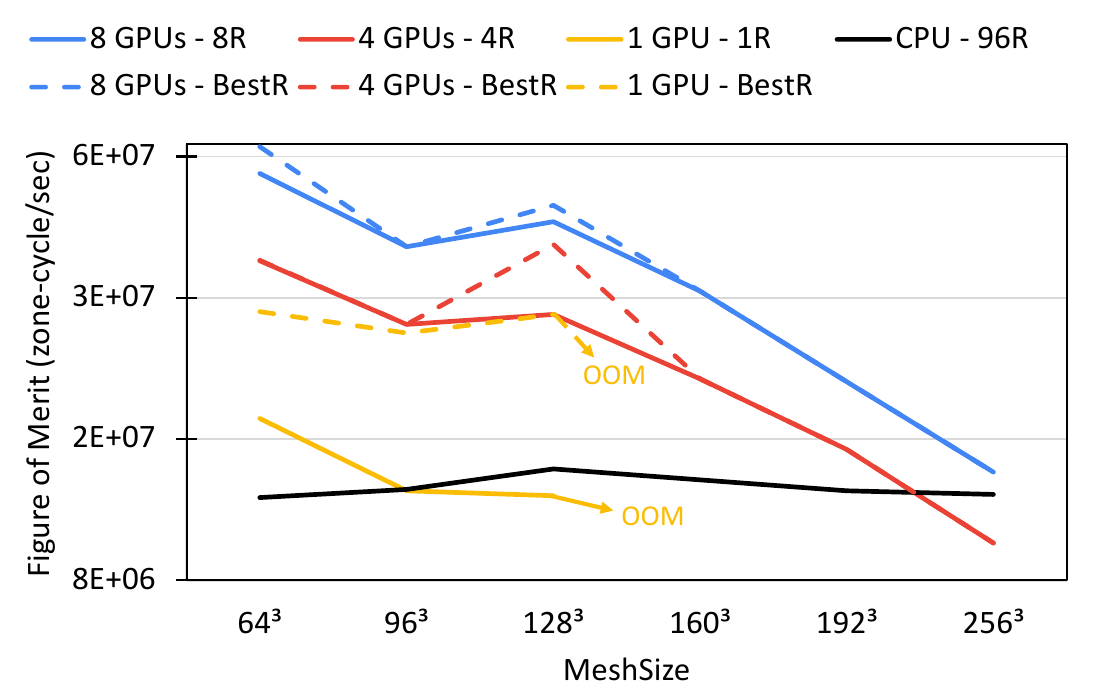}
    \caption{Performance versus Mesh Size (Static Scaling with \texttt{MeshBlockSize = 16} and \texttt{\#AMR Levels = 3})}
    \label{fig:perf_mesh_size}
\end{figure}

First, performance degrades with larger mesh size primarily due to a significant rise in the serial portion of the code, including block management and communication overhead. For instance, when the mesh size increases from 64 to 128 cells per dimension, the number of communicated cells grows by 5.9$\times$, while cell updates increase by 4.5$\times$. Consequently, the serial portion outside the Kokkos kernels increases by 5.4$\times$, and the time spent in Kokkos kernels on the GPU rises by only 2.8$\times$. Notably, GPU performance is more sensitive to mesh size than CPU performance, as GPUs are more affected by serial bottlenecks. This is, in turn, because GPU configurations have a smaller number of ranks, as detailed in Section~\ref{sec:gpurankscaling}. In contrast, CPUs efficiently support up to 96 ranks across all mesh sizes, mitigating serial overhead.

Second, larger mesh sizes can improve performance at high rank counts. For example, CPU performance with 96 ranks increases up to a mesh size of 128 despite higher serial time. This improvement occurs because smaller mesh sizes (below 128) lack enough MeshBlocks to fully utilize all 96 ranks, leading to under-utilization.

\subsection{Mesh Block Size vs Performance}

\begin{figure}[t]
    \centering
    \includegraphics[width=\linewidth]{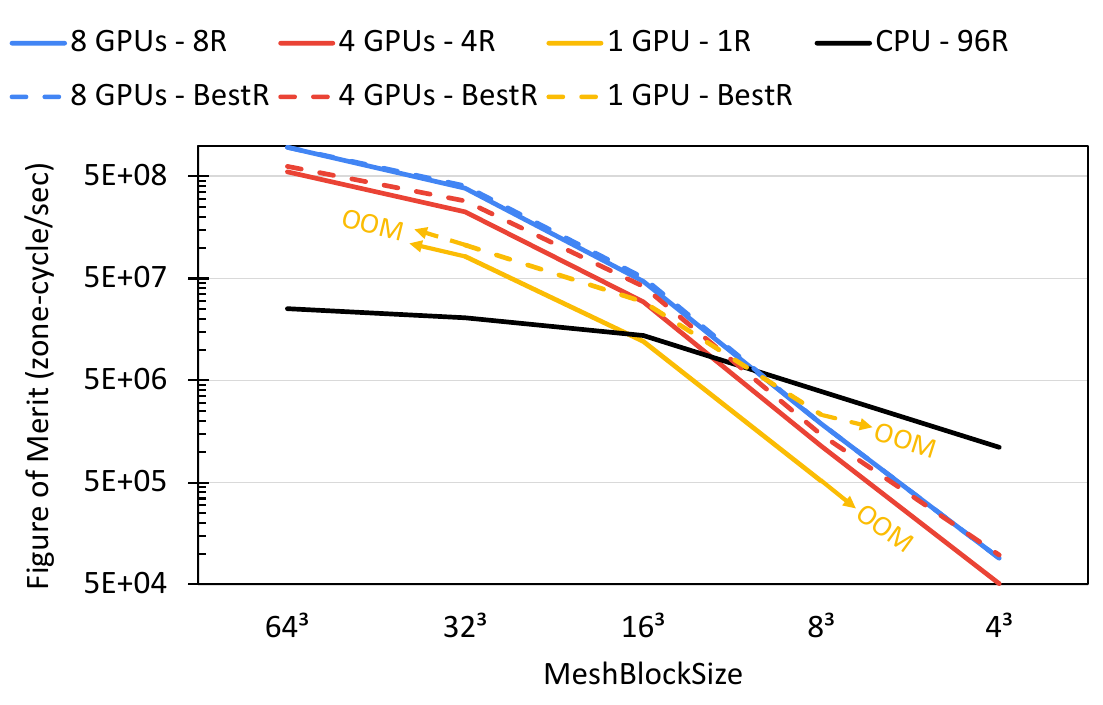}
    \caption{Performance versus \texttt{MeshBlockSize} (Mesh Size \texttt{= 128} and \texttt{\#AMR Levels = 3})}
    \label{fig:perfvsmeshblocksize}
\end{figure}

Fig.~\ref{fig:perfvsmeshblocksize} presents performance on the Y-axis as a function of \texttt{MeshBlockSize} on the X-axis. Both CPU and GPU performance decline as the mesh block size decreases; however, the decline is significantly steeper on the GPU.

The primary reason for performance degradation is the growing communication overhead relative to computation time as mesh block sizes decrease. When the \texttt{MeshBlockSize} is reduced from 32 to 16, the number of communicated cells increases by 2.1$\times$, while the total number of cell updates decreases by 5.0$\times$. This leads to an approximately 10.9$\times$ increase in the communication-to-computation ratio, significantly impacting overall performance. Reducing the block size to 8 further exacerbates the problem. For instance, the time spent in the MPI function \texttt{ReceiveBoundBufs} during the CPU execution rises by 3.6$\times$ when reducing the block size from 16 to 8.
%\todo{different configuration from the first part of this paragraph. I don't think if we need the whole sentence at all.}

% The decrease in CPU performance is because of increased communication overhead, as a larger number of smaller MeshBlocks leads to longer MPI wait times. For instance, the time spent in the MPI function \texttt{ReceiveBoundBufs} rises from 20.68 seconds at a block size of 16 to 75.32 seconds at size 8. 

%\textcolor{blue}{Additionally, smaller mesh blocks require more iterations within Kokkos kernels, increasing kernel execution time.}\todo{There will be more iterations, but each iteration will do less work. It's not clear what you mean from this.}

% When reducing the \texttt{MeshBlockSize} from 32 to 16, the number of cells communicated nearly doubles from approximately \(1.02 \times 10^{11}\) to \(2.20 \times 10^{11}\), while the number of cell updates decreases from \(5.34 \times 10^{09}\) to \(1.05 \times 10^{09}\). This results in an approximate 11$\times$ increase in the communication-to-computation ratio which adversely impacts performance.

%\textcolor{blue}{

Additionally, GPU performance declines sharply because the fraction of time spent in Kokkos kernels reduces drastically. These kernels are optimized to leverage GPU resources effectively at larger mesh block sizes. In a 1 GPU - 1 Rank configuration, total execution time increases substantially as the block size decreases: from 97.63 seconds at size 32, to 257.21 seconds at size 16, and up to 3023 seconds at size 8. 

%Correspondingly, the percentage of time spent in Kokkos kernels drops from 55.29\% to 17.97\% and then to 4.05\%, highlighting a severe loss in GPU efficiency.

%\todo{This whole paragraph don't make sense to me at all! We already explain that the serial time increases b/c of more communication. Kernel time stays the same or even reduces for both CPU and GPU. So the proportion of both CPU and GPU kernels reduce. This is not a reason for GPU performance degrading sharper than CPU. What do you mean by "These kernels are optimized to leverage GPU resources effectively at larger mesh block sizes."?}

\textbf{This analysis demonstrates that smaller mesh blocks substantially reduce GPU performance.} Consequently, at a mesh block size of 16, the performance of a single GPU is lower than that of a 96-core CPU, and at size 8, configurations using 4 GPUs perform worse than CPU.

\subsection{AMR Levels vs Performance}

\begin{figure}[t]
    \centering
    \includegraphics[width=1\linewidth]{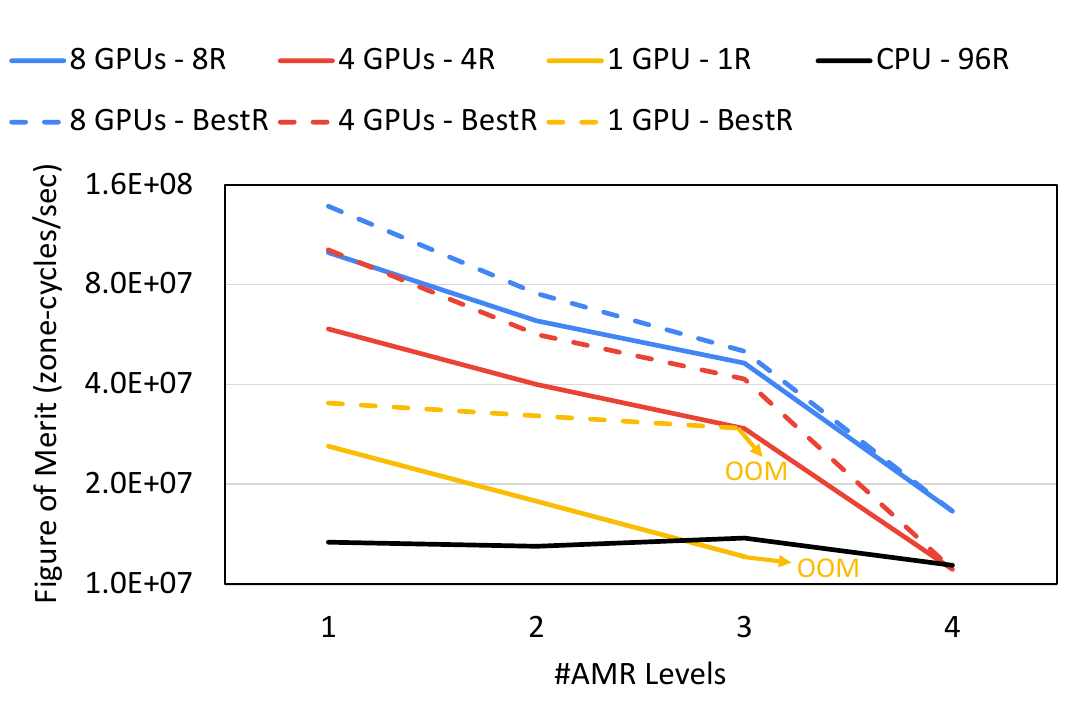}
    \caption{Performance vs \texttt{\#AMR Levels} (Mesh Size \texttt{= 128}, \texttt{MeshBlockSize = 16})}
    \label{fig:perfvsamrlevels}
\end{figure}

Fig.~\ref{fig:perfvsamrlevels} presents the effect of increasing the \texttt{\#AMR Levels} on performance, with \texttt{MeshSize} and \texttt{MeshBlockSize} fixed at 128 and 16, respectively. CPU performance remains nearly constant as \texttt{\#AMR Levels} increase, whereas GPU performance experiences a marked reduction.

This degradation on the GPU is due to a higher fraction of serial code as the \texttt{\#AMR Levels} grows. Using a 1 GPU – 1 Rank system, and compared to one AMR level, the total execution time increases by 2.1$\times$ at two levels, and by 6.0$\times$ at three levels.
% from 42.4 sec at level 1, to 90.1 sec at level 2, and then to 257.2 sec at level 3. 
Concurrently, the fraction of time spent in Kokkos kernels decreases from 31.2\% at level 1, to 23.4\% at level 2, and finally to 17.9\% at level 3.

Additionally, communication costs increase with deeper AMR hierarchies, contributing further to performance degradation. For example, with a fixed \texttt{MeshSize} of 128 and smallest block size we experiment (\texttt{MeshBlockSize}=8), compared to one level, the number of cells communicated grows by 1.4$\times$ at two levels, and 2.7$\times$ at three levels, while the number of cell updates increases by 1.2$\times$ at two levels, and 2.0$\times$ at three levels.
% from approximately \(8.07 \times 10^{10}\) at level 1, to \(1.13 \times 10^{11}\) at level 2, and up to \(2.20 \times 10^{11}\) at level 3. Meanwhile, the number of cell updates increases from \(5.24 \times 10^{8}\) to \(6.39 \times 10^{8}\), and further to \(1.05 \times 10^{9}\). 
This growing communication cost relative to computation further impacts GPU performance as AMR levels increase.

\subsection{Impact of Rank Scaling for CPU}
\begin{figure}[b]
    \centering
    \includegraphics[width=1\linewidth]{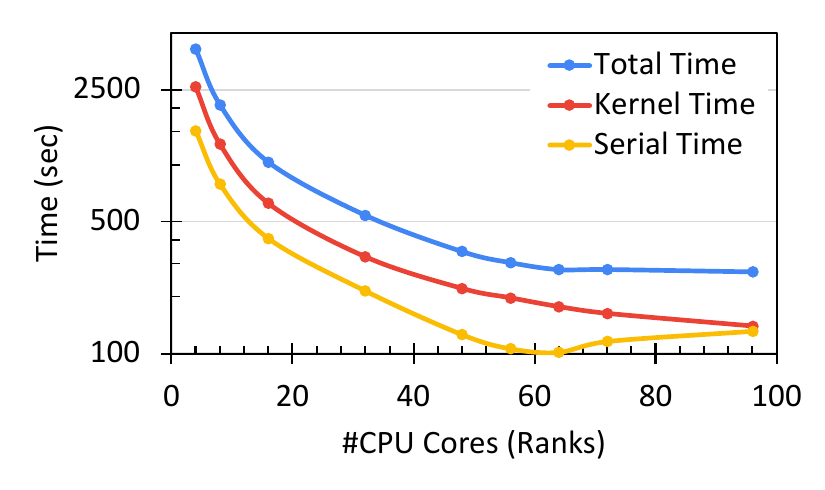}
    \caption{CPU Strong scaling - Breakdown of the Total time into kernel time and the serial portion (Mesh Size \texttt{= 128}, \texttt{MeshBlockSize = 8}, \#AMR Levels = 3)}
    \label{fig:cpu_strong_scaling}
\end{figure}

To investigate this scaling behavior further, we performed a \textit{strong scaling} experiment on CPU-only hardware, varying the core count from 4 up to 96 cores. The results are summarized in Fig.~\ref{fig:cpu_strong_scaling}. From these results, we observe the following: \textit{First}, total runtime decreases nearly ideally as core count increases from 4 to 48 cores, demonstrating strong scaling. \textit{Second}, kernel execution time also scales well up to 96 cores, reducing significantly with more cores. \textit{Third}, the serial portion of the runtime decreases steadily up to about 64 cores, after which it plateaus  indicates an irreducible serial overhead.
%, \textcolor{blue}{consistent with Amdahl's law.}
%\todo{How is this related to Amdahl's law? The serial portion indeed decreases. So it actually contradicts Amdahl's law.} 
%and slightly increases at 72 and 96 cores.
The minor increase in serial time at 72 and 96 cores suggests resource contention or synchronization overheads, potentially due to the costs of collective operations such as All-Reduce and All-Gather.
%that limit scaling efficiency at high concurrency.

%This analysis highlights the critical role of concurrency in hardware utilization efficiency. \textcolor{blue}{For GPUs, low MPI rank counts cause underutilization due to high serial overhead}\todo{we haven't discussed and shown this yet}, whereas CPUs exhibit balanced scaling but plateau in overhead reduction as concurrency grows, indicating diminishing returns at high core counts.

\subsection{Impact of Rank Scaling for GPU}~\label{sec:gpurankscaling}
We investigate GPU strong scaling with MPI ranks. Using a 1 GPU - 1 Rank configuration, the total execution time is 2782 seconds, with CPU serial time dominating at 2659 seconds and GPU kernel execution accounting for only 122 seconds for Mesh Size \texttt{= 128}, \texttt{MeshBlockSize = 8}, \#AMR Levels = 3. This highlights a strong dependence on the CPU host to handle block management. 
%\textcolor{blue}{due to insufficient host parallelism}\todo{What does this mean? It's too vague. Why do we need more ranks? to handle block scheduling, communication, and synchronization?}, which limits GPU utilization.

\begin{figure}[t]
    \centering
    \includegraphics[width=1\linewidth]{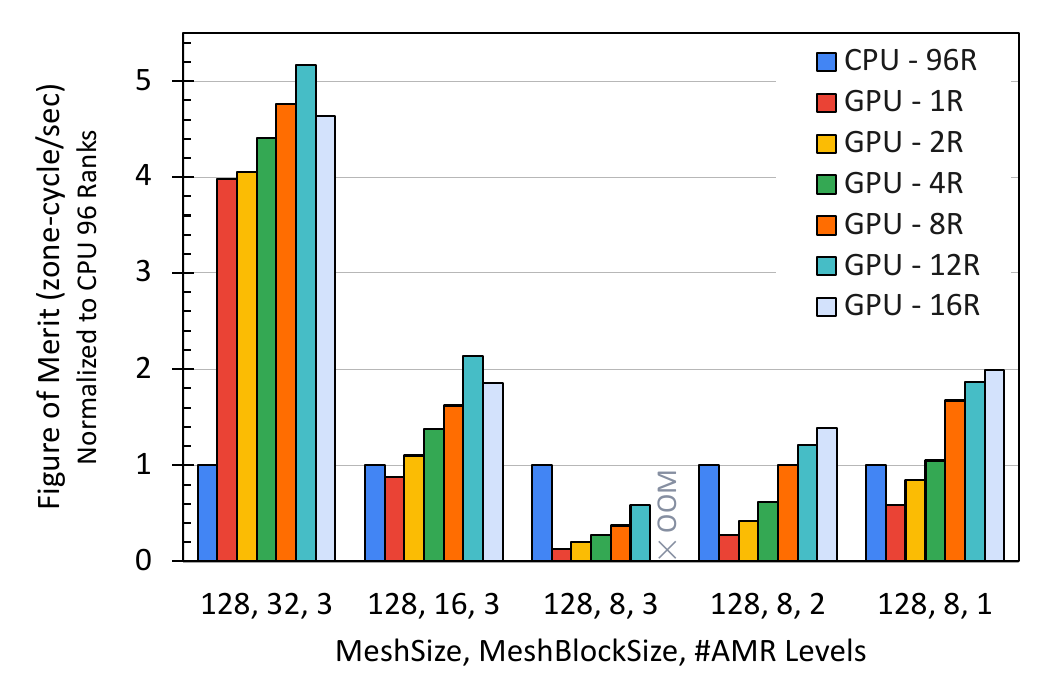}
    \caption{Effect of increasing Ranks on GPU.}
    \label{fig:ranks}
\end{figure}

To address this bottleneck, increasing the number of MPI ranks per GPU can help improve overall performance by enabling greater concurrency on the host and better overlapping of computation and communication. Fig.~\ref{fig:ranks} illustrates how performance scales as ranks per GPU increase in different AMR configurations, showing substantial gains up to a certain point.

We observe that the best performance on single-GPU systems is achieved at around 12 ranks per GPU. Beyond this point, increasing the rank count further leads to performance degradation. This drop is partly due to overheads from collective communication operations such as All-Gather and All-Reduce, which become more expensive with higher rank counts. 
% \todo{\textbf{Does this Make Sense?} This drop is also driven by increasing metadata overhead, worsening load imbalance across ranks, and excessive ghost zone communication relative to useful work. As MeshBlock granularity shrinks and rank count grows, the per-rank computation-to-communication ratio decreases significantly, especially in routines like \texttt{SendBoundBufs}, \texttt{EstimateTimestep}, and \texttt{RedistributeAndRefineMeshBlocks}.} 
In contrast, CPU runs do not show a similar drop in overall FOM with more core counts. Kernel execution times on CPUs continue to decrease steadily with additional cores, and higher serial overhead is outweighed by parallel performance gains (Fig.~\ref{fig:cpu_strong_scaling}). 
%This trend is visible in the strong scaling data presented in Figure~\ref{fig:cpu_strong_scaling}. 

\begin{figure}[t]
    \centering
    \includegraphics[width=1\linewidth]{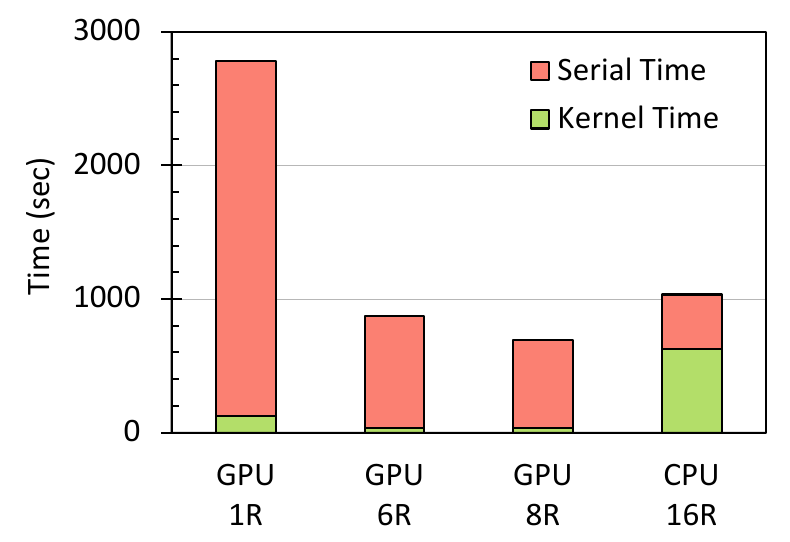}
    \caption{Execution time breakdown for MeshSize=128, \texttt{MeshBlockSize=8}, and \texttt{\#AMR Levels=3}.}
     % across hardware configurations 
     % The bars show total execution time and breakdown to kernel portion (Kokkos time), and serial portion (time outside Kokkos kernels).
    \label{fig:kernel_timing_breakdown}
\end{figure}

Fig.~\ref{fig:kernel_timing_breakdown} presents the breakdown of execution time between Kokkos kernels (offloaded to GPU) and the serial portion (runs on CPU host).
% The X-axis shows hardware setups: GPUs with 1, 6, and 8 MPI ranks.  
GPUs with low concurrency (\textit{e.g.,} 1 rank) spend a large fraction of total time outside Kokkos kernels %\textcolor{blue}
{(approximately 2659 seconds)} for Mesh Size \texttt{= 128}, \texttt{MeshBlockSize = 8} and \texttt{\#AMR Levels = 3}, highlighting significant serial overhead that limits GPU utilization. 

%CPUs show decreasing total execution times with increasing ranks, indicating effective scaling. Kernel time represents a significant portion of total time, especially at lower ranks, illustrating a more balanced division between kernel and non-kernel computation.

%The Y-axis shows total execution time in seconds, separated into total kernel time (Kokkos kernels) and time outside these kernels.

\begin{figure}[t]
\centering
\includegraphics[width=1\linewidth]{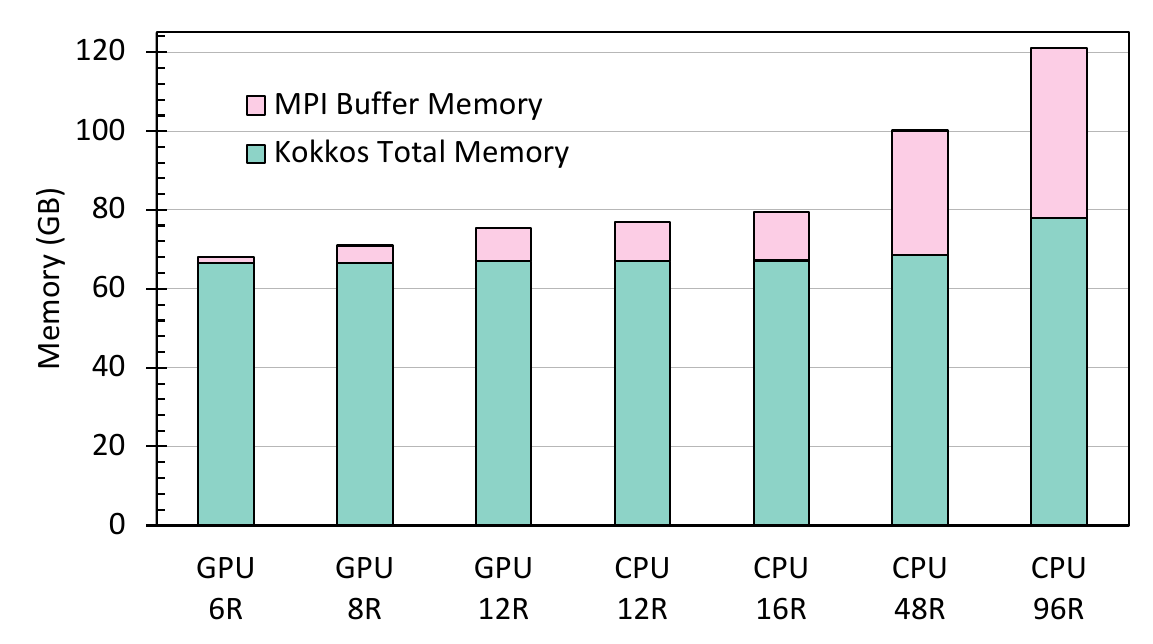}
\caption{Memory usage breakdown by Kokkos (green) and MPI communication buffers + Open MPI driver (pink) for MeshSize=128, MeshBlockSize=8, and \#AMR Levels=3. GPU 6R, 8R, 12R configurations only show GPU device memory usage. }
\label{fig:memory}
\end{figure}

However, scaling ranks per GPU is limited by memory capacity. For smaller mesh blocks, increasing ranks beyond a certain threshold leads to out-of-memory (OOM) errors, limiting scalability. The GPU device memory is 80 GB, which is substantially lower than the CPU memory capacity of 1 TB. 
% \textcolor{red}{
This trade-off underscores the importance of increasing rank parallelism with available GPU memory resources.
% } 
For example, the memory consumed by 1 GPU with 12 ranks (75.5 GB) for Mesh Size \texttt{= 128}, \texttt{MeshBlockSize = 8} and \texttt{\#AMR Levels = 3} reaches near GPU's HBM capacity. 
%We can run maximum of 16 ranks with the GPU.

% \textcolor{blue}{
To analyze \textit{GPU device memory usage}, we first use \texttt{nvidia-smi} to profile total memory consumption. To attribute this memory usage to specific code regions, we employ Kokkos Tools~\cite{kokkos_tools} and the Nvidia Nsight System Profiler to trace memory allocations and deallocations along with their associated call stacks. Our analysis of trace files reveals two sources of GPU device memory consumption: (1) allocations by Parthenon and Kokkos for mesh data, and (2) memory used by MPI communication buffers and the Open MPI driver.
% }

Fig.~\ref{fig:memory} shows that the increase in memory consumption with more MPI ranks is driven significantly by overheads such as MPI communication buffers and Open MPI drivers~\footnote{Open MPI Driver suffers from a bug~\cite{openmpi_bug} Inter-Process Communication (IPC), which leads to memory leakage. We expect that resolving this bug reduces the memory consumption.} 
% \textcolor{red}{
than by Kokkos-managed allocations.
% } 
Additionally, Kokkos-managed allocations are a significant fraction of the memory footprint and nearly constant for the same problem size. We propose code optimizations in Section~\ref{subsec:mem_opt} to reduce Kokkos-managed memory and enable more ranks. 
%\textcolor{blue}{As the number of ranks increases, total memory usage grows noticeably, with MPI-related buffers accounting for an increasingly larger fraction at higher ranks.}\todo{redundant. Remove it?}

Overall, the primary GPU performance bottleneck arises from the serial portions of the code that are executed by host CPU ranks. Exploiting additional CPU cores alongside GPUs and increasing ranks per GPU presents a promising approach to mitigating this limitation and enhancing GPU utilization. From our CPU rank scaling studies (Fig.~\ref{fig:cpu_strong_scaling}), we expect the performance to flatten out after 48 ranks.

\section{Multi-Node Discussion}

We used the same hardware configurations as detailed in Tables~\ref{tab:cpu_specs} (96 Sapphire Rapids cores) and~\ref{tab:gpu_specs} (8 H100s), but deployed two nodes for each platform. All results in this section use one MPI rank per GPU for GPU experiments and one rank per core for CPU experiments. CPUs demonstrate strong scaling with additional nodes, whereas GPUs show notably weaker scaling. For \texttt{MeshSize = 128}, \texttt{MeshBlockSize = 8}, and \texttt{\#AMR Levels = 3}, the performance improvement (ratio of two nodes to one node) was 1.63$\times$ for CPUs but 1.51$\times$ for GPUs. At a larger \texttt{MeshBlockSize = 16}, CPUs scaled even better with a 1.85$\times$ improvement, while GPUs decrease slightly (0.95$\times$).

The performance reduction when using smaller mesh blocks remains far more severe for GPUs than for CPUs, even in multi-node settings. For both \texttt{MeshSize = 128} and \texttt{\#AMR Levels = 3}, the performance drop from large to small mesh blocks (\texttt{MeshBlockSize} 32 to 8) across two CPU nodes was 5.88$\times$, whereas for two GPU nodes it was a dramatic 90.77$\times$. At a larger \texttt{MeshSize = 256}, the drop remained stable for CPUs (5.73$\times$) but increased for GPUs to 207.83$\times$. Deeper AMR levels also hurt GPUs more: for \texttt{MeshSize = 256} and \texttt{MeshBlockSize = 16}, the performance drop from 1 to 3 levels for two CPU nodes is 1.22$\times$, while two GPU nodes see a much larger drop of 3.92$\times$.

Together, these results reinforce our single-node findings: GPUs are more vulnerable to fine-grained AMR configurations, and multi-node communication exacerbates their underutilization. We stress that these results represent an initial characterization, and more nodes and alternative rank mappings will be necessary for broader claims regarding exascale AMR performance.

\section{Timing Analysis} \label{sec:topdown}

\begin{figure}[t]
    \centering
    \includegraphics[width=\linewidth]{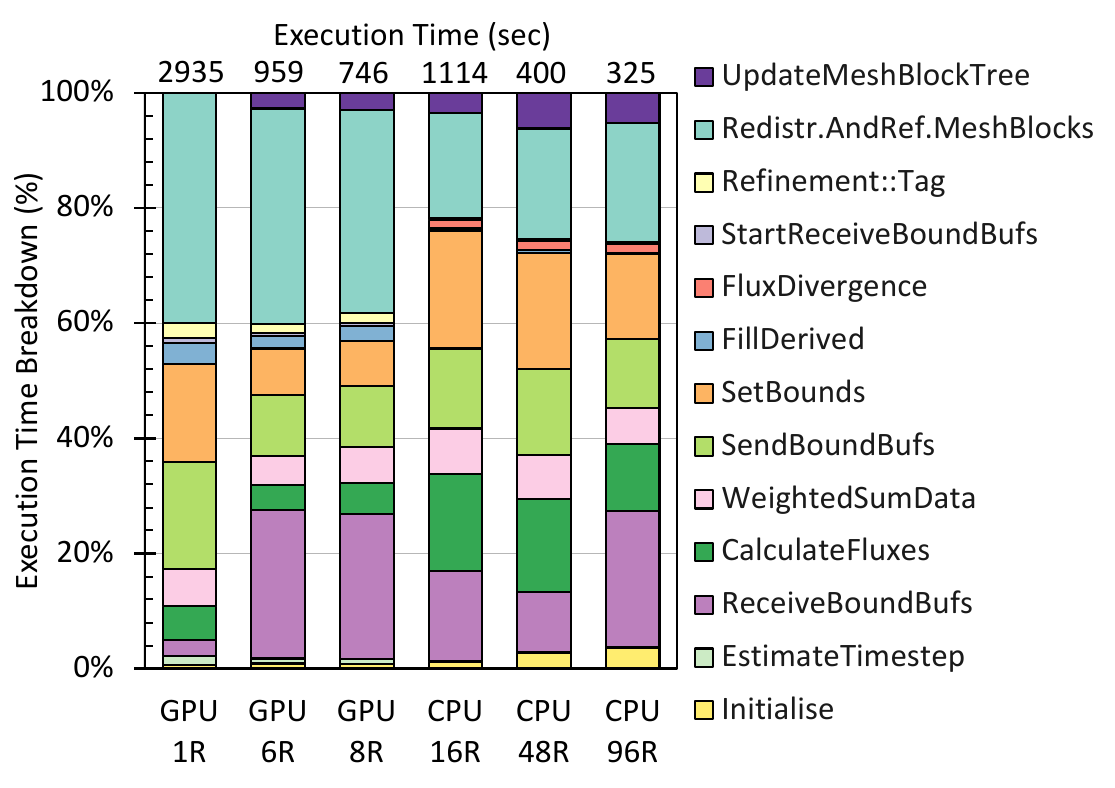}
    \caption{Execution time breakdown of key functions across different hardware configurations for MeshSize=128, \texttt{MeshBlockSize=8}, and \texttt{\#AMR Levels=3}.}
    \label{fig:stackedbarchart}
\end{figure}

To understand performance bottlenecks and execution characteristics within Parthenon, we perform a top-down and hotspot analysis focused on the core timestep execution loop. 
%\subsection{Timing Breakdown}
Fig.~\ref{fig:stackedbarchart} presents a stacked bar chart showing the breakdown of runtime across key functions in the Parthenon timestep loop. The experiment is conducted using a mesh size of 128, a mesh block size of 8, and 3 AMR levels. The X-axis captures different hardware configurations—GPU executions with 1, 6, and 8 ranks, and CPU executions with 16, 48, and 96 ranks. Each bar is normalized to 100\% and shows the proportion of total time spent in each function. The absolute total runtime in seconds for each configuration is shown above the respective bar for reference.

Across all configurations, we observe that GPU executions with fewer ranks (\textit{e.g.}, GPU -- 1R) spend a significant amount of time in serial or CPU-resident portions of the code, particularly in \texttt{RedistributeAndRefineMeshBlocks}, \texttt{SendBoundBufs}, and \texttt{SetBounds}. These trends corroborate earlier observations on the high overhead of serial code paths when GPU concurrency is low.

As we increase the number of MPI ranks on the GPU, many of these overheads decrease significantly. For example, time spent in \texttt{RedistributeAndRefineMeshBlocks} drops from over 1100 seconds in the 1-rank GPU run to just 263 seconds in the 8-rank GPU case. A similar trend is seen for \texttt{SendBoundBufs} and \texttt{SetBounds}, which are also key contributors to communication and packing overheads.

CPU configurations show more balanced distributions, with lower overall time spent in communication routines. Functions like \texttt{CalculateFluxes} and \texttt{WeightedSumData} dominate early on (16 ranks), but shrink as we scale to 96 cores, indicating good strong scaling. Notably, \texttt{ReceiveBoundBufs} and \texttt{SendBoundBufs} remain steady contributors even at high CPU concurrency, underlining the persistent cost of ghost zone communication.

Functions such as \texttt{Refinement::Tag} and \texttt{UpdateMeshBlockTree} appear only as small slices in the total time but are highly serial and dominant in GPU runs with fewer ranks. This reinforces the need to optimize serial phases to improve GPU utilization.
\begin{figure*}[h]
    \centering
    \includegraphics[width=1\linewidth]{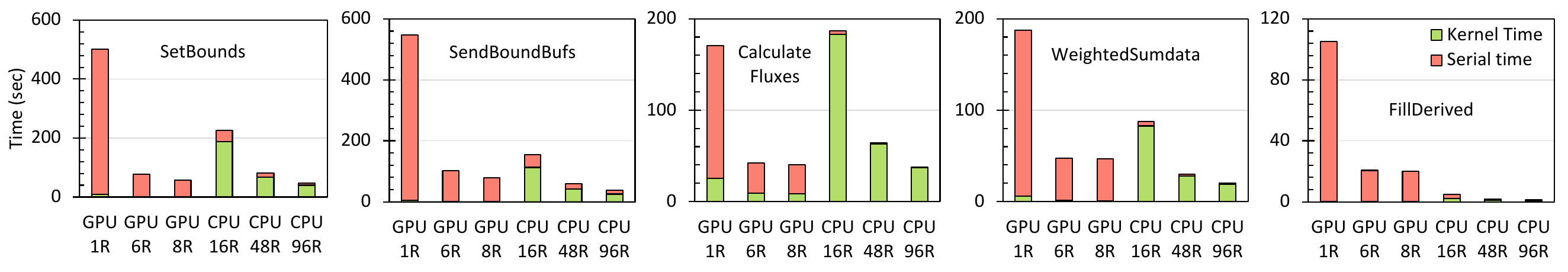}
    \caption{Execution time breakdown for key Parthenon functions across different hardware configurations, using a MeshSize=128, \texttt{MeshBlockSize=8}, and \texttt{\#AMR Levels=3}.}
     % Each subplot compares total execution time (blue bars) with GPU-offloaded kernel time (red bars) for five critical functions: \texttt{SetBounds}, \texttt{SendBoundBufs}, \texttt{CalculateFluxes}, \texttt{WeightedSumData}, and \texttt{FillDerived}
    \label{fig:functions}
\end{figure*}

Fig.~\ref{fig:functions} presents a detailed timing breakdown for critical functions within the Parthenon timestep. The execution time is broken down into the serial and GPU-offloadable kernel times. The tests use a mesh size of 128, a mesh block size of 8, and 3 levels of AMR.
% Each subfigure compares the total execution time versus GPU-offloaded kernel time (where applicable) for the following key functions: \texttt{SetBounds}, \texttt{SendBoundBufs}, \texttt{CalculateFluxes}, \texttt{WeightedSumData}, and \texttt{FillDerived}.
Across all functions, GPU executions with a single MPI rank exhibit a substantial gap between serial and kernel times, indicating that serial or CPU-resident operations dominate the runtime. 

Overall, timing analysis shows that communication and load balancing operations present significant bottlenecks
% at lower concurrency, especially
in GPU runs with fewer ranks. Increasing the number of ranks effectively reduces these overheads by enhancing parallelism and better utilizing hardware resources. This analysis motivates further optimization of communication and load balancing mechanisms to improve scaling efficiency in high concurrency environments.

%\subsection{Timing Breakdown for Key Functions} 

%% this seems repetative not providing new insights so commenting this out.
\section{Microarchitecture Characterization}

\subsection{GPU Microarchitecture Analysis}
\label{sec:microarch}

While Section~\ref{sec:performance} identifies serial code as a major bottleneck for GPU performance, Table~\ref{tab:gpumicroarch} further analyzes GPU microarchitectural characteristics \textit{during active cycles}. This analysis focuses on 10 most time-consuming kernels using a mesh size of 128, block sizes of 32 and 16 (B32 and B16), and three levels of AMR. Total is reported as a weighted average based on kernel durations. Notably, the \texttt{CalculateFluxes} kernel alone accounts for 41.0\% and 36.9\% of total kernel time in B32 and B16, respectively.

Our analysis shows that SM utilization is primarily limited by SM occupancy, which is in turn constrained by the high register usage of CUDA kernels. SM occupancy, defined as the ratio of active warps to the maximum number of supported warps, is critical for hiding memory latency. However, large register, shared memory, and barrier requirements, as well as suboptimal CUDA grid/block configurations can all limit occupancy. Between these, our code analysis reveals that a significant register requirement is the main reason for limited occupancy in the evaluated kernels. In particular, \texttt{CalculateFluxes} uses over 100 registers per thread, limiting active warps per SM to only four.

\begin{table}[t]
    \caption{GPU Microarchitecture Analysis}
    \scriptsize
    \renewcommand\arraystretch{1}
    \scalebox{1}{
        \setlength{\tabcolsep}{1.0pt} % Default value: 6pt
        \renewcommand{\arraystretch}{1.2}
        \begin{tabular}{|c|c|c|c|c|c|c|c|c|c|c|c|c|}
            \hline
            \multirow{3}{*}{Kernel} & \multicolumn{2}{|c|}{Duration*} & \multicolumn{2}{c|}{SM Util.} &  \multicolumn{2}{c|}{SM Occ.} & \multicolumn{2}{c|}{Warp Util.} & \multicolumn{2}{c|}{BW Util.} & \multicolumn{2}{c|}{Arith. Int.} \\
            & \multicolumn{2}{|c|}{(ms)} & \multicolumn{2}{c|}{(\%)} &  \multicolumn{2}{c|}{(\%)} & \multicolumn{2}{c|}{(\%)} & \multicolumn{2}{c|}{(\%)} & \multicolumn{2}{c|}{(FLOPs/B)} \\
            \cline{2-13}
            & B32 & B16 & B32 & B16 & B32 & B16 & B32 & B16 & B32 & B16 & B32 & B16 \\
            \hline
            \hline
            CalculateFluxes & 135.0 & 94.9 & 32.3 & 27.9 & 24.1 & 24.2 & 94.1 & 67.6 & 18.5 & 11.2 & 4.3 & 3.4 \\
            \hline
            FirstDerivative & 63.8 & 51.0 & 2.5 & 2.2 & 52.3 & 52.5 & 95.9 & 94.4 & 0.1 & 0.1 & 14.5 & 18.7 \\
            \hline
            MassHistory & 35.6 & 35.7 & 5.6 & 4.0 & 24.2 & 24.1 & 100.0 & 50.0 & 1.8 & 0.8 & 3.1 & 2.6 \\
            \hline
            WeightedSumData & 27.1 & 21.1 & 69.1 & 54.5 & 92.7 & 94.2 & 94.8 & 100.0 & 50.2 & 39.6 & 0.3 & 0.3 \\
            \hline
            SendBoundBufs & 23.7 & 15.9 & 5.5 & 11.3 & 95.7 & 97.9 & 94.4 & 84.3 & 28.5 & 34.2 & 0.0 & 0.0 \\
            \hline
            SetBounds & 20.9 & 20.9 & 12.4 & 14.3 & 51.5 & 50.4 & 94.2 & 88.4 & 22.2 & 21.5 & 0.1 & 0.1 \\
            \hline
            FluxDivergence & 11.9 & 4.9 & 48.5 & 41.6 & 94.5 & 97.5 & 95.0 & 100.0 & 51.2 & 51.8 & 0.6 & 0.5 \\
            \hline
            Est.Time.Mesh & 6.5 & 8.3 & 3.7 & 2.9 & 24.2 & 24.1 & 94.7 & 50.1 & 3.3 & 1.6 & 1.7 & 1.4 \\
            \hline
            Prolong.Restr.Loop & 3.3 & 3.1 & 24.8 & 29.7 & 54.9 & 66.3 & 94.9 & 93.4 & 56.9 & 54.5 & 0.3 & 0.5 \\
            \hline
            CalculateDerived & 1.3 & 0.9 & 39.2 & 46.8 & 36.9 & 41.9 & 94.3 & 74.4 & 54.1 & 37.4 & 0.1 & 0.1 \\
            \hline
            \hline
            Total & 329.1 & 256.7 & 23.4 & 19.1 & 45.0 & 44.2 & 95.3 & 76.3 & 18.1 & 13.2 & 5.0 & 5.4 \\
            \hline
            \multicolumn{11}{l}{*CUDA kernel time during a single cycle of simulation.}
        \end{tabular}
    }
    \label{tab:gpumicroarch}
\end{table}

Warp utilization, measured as the ratio of active threads per warp instruction, is generally high. However, the \texttt{CalculateFluxes} kernel suffers from two major inefficiencies revealed through PTX code inspection:
\textit{First}, CUDA blocks are over-provisioned at 128 threads (four warps) per block, yet only one warp performs useful computations. The remaining warps execute unnecessary indexing, data access, and control flow instructions before exiting. These ineffective instructions account for 78\% of total warp instructions per mesh block. \textit{Second}, the effective warp experiences significant control divergence, especially with smaller mesh blocks. Each warp computes fluxes along a single mesh block dimension; when the mesh block size is 16, half of the warp threads are deprecated off, reducing warp utilization. These inefficiencies suggest that aligning 3D CUDA block sizes with the mesh block dimensions could reduce idle threads and control divergence.

Although the CUDA kernels are memory bandwidth-bound due to their low arithmetic intensity (FLOPs/Byte), DRAM bandwidth utilization remains low. The low arithmetic intensity arises from 64-bit floating-point operations, which require significant data. Additionally, many kernels only perform data copies with an arithmetic intensity below one. While the H100 GPU provides 10.1 FLOPs/Byte operational density~\footnote{H100's operational intensity is calculated using 34 TFLOPS FP64 throughput and 3.35 TB/s HBM bandwidth~\cite{nvidiah100}.}, the analyzed kernels utilize only 5.0–5.4 FLOPs/Byte on average. So, we expect memory bandwidth to be yet another key bottleneck for SM utilization. However, we observe the limited memory bandwidth utilization due to the sparse memory access pattern to the mesh blocks in the memory, which lack spatial locality.

\subsection{CPU Instruction Analysis}

\begin{figure}[t]
    \centering
    \includegraphics[width=\linewidth]{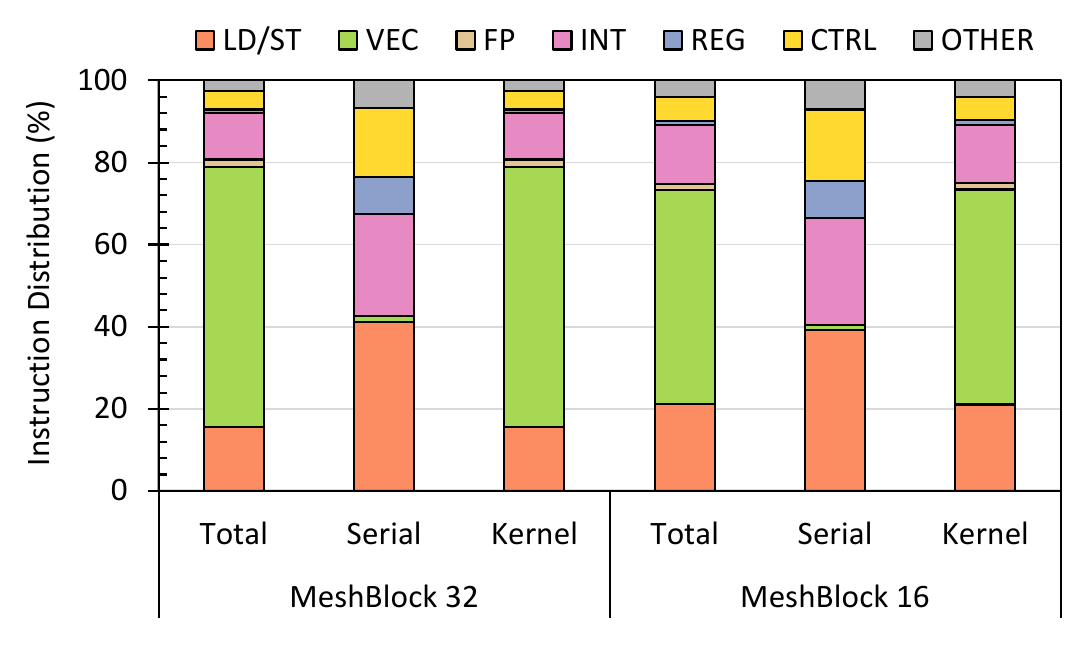}
    % \vspace{-6mm}
    \caption{CPU instruction opcode distribution for MeshSize=128, MeshBlockSize=16 and 32, and \texttt{\#AMR Levels=3}.}
    \label{fig:instr_distr}
\end{figure}

We use MICA~\cite{MICA}, an Intel Pin tool, to analyze the instruction opcode distribution on the CPU. Figure~\ref{fig:instr_distr} shows the distribution for a mesh size of 128, mesh block sizes of 32 and 16, three levels of AMR, and 16 MPI ranks. The breakdown includes \textit{Total} (entire CPU execution), \textit{Serial} (code shared by CPU and GPU runs), and \textit{Kernel} (real computation). 
Our analysis reveals several key insights. \textit{First}, vector opcodes dominate both the \textit{Total} and \textit{Kernel} instructions. SIMD instructions are mainly responsible for data-parallel arithmetic computations. \textit{Second}, \textit{Kernel} instructions constitute over 99\% of total executed instructions. These two findings assert the efficiency of CPU implementation. \textit{Third}, load and store operations account for 39--41\% of \textit{Serial} execution, due to managing block-sparse data structures. \textit{Lastly}, the share of vector instructions in data-parallel kernels decreases from 63\% to 52\% when reducing the mesh block size from 32 to 16, highlighting the improved computational efficiency of larger blocks.

%Both \texttt{SendBoundBufs} and \texttt{SetBoundBufs} call this function, further compounding its overhead in communication-intensive phases.

%Overall, these serial components introduce fixed overheads that have a greater impact on performance in low-rank runs, where limited parallelism prevents these costs from being spread out effectively. While these functions perform better as the number of ranks increases, their absolute cost remains significant when concurrency is low. The combination of string-based lookups, sorting/randomization steps, memory management in buffer caching, and scalar heuristic evaluations restricts scalability, emphasizing the need for targeted optimization in these areas.

\section{Recommendations for Scalability}

\subsection{Improving Performance of Serial Portions}
Analysis of the serial portions of Parthenon's code reveals critical bottlenecks, especially at low MPI rank counts or when using fewer cores. These CPU-bound tasks include variable management, communication setup, buffer initialization, and refinement tagging.

\textbf{String-based Indexing and Variable Lookup:} Core functions such as \texttt{CalculateFluxes}, \texttt{WeightedSumData}, and \texttt{FillDerived} rely heavily on \texttt{GetVariablesByFlag} for extracting variables using metadata flags. This mechanism depends on string-based indexing to identify variables over a scalar loop. The flexibility of this approach comes at the cost of repeated string comparisons and hashing during every lookup, causing substantial overhead.

Runtime string-based indexing could be replaced by a compile-time or integer-based indexing scheme. Employing a centralized mapping from variable names to integer identifiers may reduce the overhead of repeated string operations. Parallelizing variable lookups with OpenMP is possible, but limited, since extraction involves scalar metadata operations that are not inherently parallelizable.

\textbf{Initialization Overhead in Buffer Cache Setup:} The \texttt{InitializeBufferCache} function involves iterating over all mesh boundaries to prepare communication buffers. A key contributor to serial inefficiency here is the sorting and randomization of boundary keys.
%, specifically the process of sorting keys by receiver indices and then shuffling them. 
This function is invoked internally by \texttt{SendBoundBufs}, making it a recurring cost during every communication phase.

The sorting and randomization of boundary keys could be optimized. While randomization may sometimes improve runtime behavior by improving load balancing, these operations nonetheless add overhead. This represents a tradeoff between potential runtime improvements and added serial overhead. Parallel sorting algorithms may offer gains.

\textbf{Metadata Filling and Buffer Cache Rebuilding:} During \texttt{RebuildBufferCache}, complex data structures such as ViewsOfViews are allocated and populated with metadata like buffer sizes, restriction and prolongation flags, and other boundary information. This stage requires sequential iteration over boundaries, with costly memory allocations and data transfers, including host-to-device copies.

Batching memory allocations using Software memory pools
% and enabling asynchronous or overlapped host-device memory transfers 
% For Akash - Is batch memory allocations the same as using memory pools, ie allocate a big chunk of memory, then have the application manage the smaller allocations from the pool?
could reduce serialization delays. Parallel iteration over boundaries using OpenMP is feasible if thread safety can be ensured during metadata population, potentially improving performance in this phase. For context, the \texttt{RebuildBufferCache} function accounts for approximately 13.3\% of total runtime in a 1 GPU -- 1 Rank configuration with Mesh Size 128, \texttt{MeshBlockSize} 16, and 3 AMR levels.

\textbf{Refinement Tagging via Scalar Loops:} The decision to refine or derefine mesh blocks is made through scalar loops over all mesh blocks and their packages, invoking heuristic functions such as \texttt{CheckAllRefinement}. These heuristics aggregate refinement tags by sequentially querying each package, and involve conditional logic. The serial nature of these loops, coupled with the dependency on user-defined refinement criteria, constrains parallelism and can be optimised further.

\subsection{Reducing Memory Footprint for More Ranks}~\label{subsec:mem_opt}

%Kernel fusion is an effective optimization that reduces the memory footprint of intermediate variables and improves performance by minimizing kernel launch overheads and combining sequential computational stages into a single kernel. For instance, in \texttt{CalculateFlux}, this technique can merge the reconstruction of state variables at cell faces with flux calculations from Riemann solvers. 

%By eliminating intermediate memory writes between these stages, the fused kernel reduces launch overhead, improves data locality, and maintains correctness by explicitly preserving data dependencies. This fusion also allows shrinking the amount of required scratch memory—from a size proportional to the entire mesh down to a local size multiplied only by the number of active thread teams—thereby further optimizing memory usage.

Restructuring Kokkos kernels can also lead to substantial memory footprint reduction. Current three-dimensional Kokkos kernels, such as \texttt{CalculateFlux}, only launch CUDA kernels for the innermost dimension. For a smaller MeshBlock size this leads to poor SM utilization as we discuss in Section~\ref{sec:microarch}, degrading GPU performance. To address this problem, kernels can be restructured to use two or three dimensions. More importantly, this restructuring significantly reduces the memory footprint required for \textit{ auxiliary intermediate variables and temporary storage}.  
%For three-dimensional Kokkos kernels such as \texttt{CalculateFlux}, smaller meshblock sizes limit iteration counts along spatial dimensions, reducing total parallel workload per kernel and degrading GPU performance. To address this, kernels can be restructured into effectively two-dimensional loops using loop unrolling techniques, increasing computational intensity per invocation and improving execution efficiency.
%Importantly, this optimization does not alter the base mesh resolution or communication logic; it solely reduces the memory footprint required for auxiliary intermediate variables and temporary storage. 
Memory usage of auxiliary variables before and after this optimization can be modeled as:
\[
\#\text{MeshBlocks} \times B \times 6 \times (nx1 + 2 \times ng)^{\text{dimension}} \times (3 + num\_scalar)
\]
\[
\#\text{ThreadBlocks} \times B \times 6 \times (nx1 + 2 \times ng)^{d} \times (3 + num\_scalar)
\]

Here, the factor 6 accounts for three spatial directions and two sides per direction, the exponent corresponds to the spatial dimensionality (3D) pre-optimization, and \(d\) denotes the reduced dimension (e.g., 2 for a 2D loop) post-optimization. Here, \(nx1\) denotes the MeshBlock size per spatial dimension. Additionally, \(B\) is the number of bytes per variable, \(ng\) is the number of ghost cells (typically 4 for WENO5), and \(3 + num\_scalar\) corresponds to the fixed components (primary conserved quantities) plus extra passive scalars or user-defined quantities.

This reduction stems from two factors: (1) The cubic term reduces to a square (or lower dimensional) term by eliminating the need for full 3D intermediate buffers per mesh block, instead processing smaller 2D data segments within a single kernel. (2) The shift from the number of mesh blocks to the number of GPU thread blocks reflects improved memory sharing and reuse across warps, lowering overall memory allocation.

For example, in the Burgers benchmark, where \(num\_scalar=8\), \(nx1=8\), \#ThreadBlocks = 1024 (typical for an H100 GPU), \(B=8\) bytes per variable, and \(ng=4\), memory usage can be reduced from 8.858 GB before optimization to about 0.138 GB after, representing a substantial reduction in memory footprint.

\section{Related Works}

While this work focuses on a synthetic benchmark to systematically characterize Parthenon's behavior, many real-world scientific applications are built on Parthenon, including \textit{Riot}~\cite{parthenon} for multi-material radiation hydrodynamics, \textit{AthenaPK}~\cite{athenapk} for astrophysical turbulence, \textit{Artemis}~\cite{artemis} for planet formation, and \textit{KHARMA}~\cite{kharma,kharma2} for accretion flows around supermassive black holes. These codes inherit Parthenon's mesh block decomposition and hierarchical AMR, and therefore encounter similar trade-offs and GPU bottlenecks explored in this paper.

Although our analysis centers on Parthenon, other GPU-accelerated AMR frameworks, such as AMReX and Uintah, adopt different strategies for mesh management while facing similar performance challenges.
AMReX~\cite{AMReX} is a widely-used GPU-accelerated AMR framework that organizes computation through a hierarchy of grids at each refinement level, similar to Parthenon's mesh block approach. Uintah~\cite{uintah}, designed for scalable multiphysics and hazard analysis, employs an octree-based AMR structure. 
Like Parthenon, these frameworks must address issues such as kernel reshaping, MPI rank scaling, and ghost-zone communication overhead on heterogeneous architectures.

% Many of the performance challenges we identify—such as kernel reshaping, MPI rank scaling, and ghost-zone communication overhead—are relevant and applicable to these and other AMR frameworks operating on heterogeneous architectures.

% The most prominent real-world scientific application\cite{realworld} built on Parthenon is \textit{Riot}\cite{parthenon}, a multi-material radiation hydrodynamics code for high-energy density physics experiments. Other production applications include \textit{AthenaPK} \cite{athenapk} for astrophysical turbulence, \textit{Artemis} \cite{artemis} for planet formation, and \textit{KHARMA} \cite{kharma, kharma2} for accretion onto supermassive black holes. These applications are expected to encounter the same mesh block–level trade-offs and GPU execution bottlenecks that we profile in this work.

% Related GPU-accelerated AMR efforts also provide context for our study. For example, many of our findings, such as kernel reshaping, rank scaling, and ghost-zone communication, are relevant to other AMR frameworks. For example, AMReX \cite{AMReX} divides patches into "components" that play a similar role to Parthenon’s MeshBlocks, and our findings on MeshBlockSize translate naturally. Uintah \cite{uintah} (for example) is octree-based, and so the lessons learned in this study should translate directly.

\section{Conclusion} \label{sec:conclusion}

% Hero-class scientific simulations on exascale computing platforms demand high accuracy while managing extreme compute and memory requirements. Adaptive Mesh Refinement (AMR) is a critical technique enabling such simulations by dynamically focusing computational effort where it is most needed. Parthenon is a state-of-the-art, block-structured AMR framework designed for modern heterogeneous CPU-GPU architectures.

% This work provides a comprehensive performance characterization of the Parthenon. Our analysis shows that larger mesh sizes, smaller mesh block sizes, and deeper AMR levels negatively impact GPU performance. These effects arise primarily from increased serial overhead, higher communication costs, and inefficient kernel utilization due to fine-grained mesh partitioning.

Adaptive Mesh Refinement (AMR) reduces compute and memory demands in hero-class HPC simulations. We analyze the performance of Parthenon-VIBE on CPU and GPU systems and find that larger mesh sizes, smaller mesh blocks, and deeper AMR levels degrade performance due to significant serial overhead from block management and communication. While adding more MPI ranks can help reduce the serial overhead, this is limited by GPU memory capacity. We perform GPU microarchitectural analysis and show low SM occupancy and poor HBM bandwidth utilization. Finally, we suggest effective 
% code optimizations to reduce the overhead imposed by the serial portion of the code. 
software optimizations targeting serial portion of the code, such as reducing string-based metadata lookups, improving buffer cache management, and restructuring Kokkos kernels to reduce memory footprint.

% that GPU occupancy is constrained primarily by high register usage and suboptimal CUDA configurations, while memory bandwidth remains underutilized due to sparse access patterns. On the CPU side, vectorized execution dominates compute kernels, but serial operations managing block-sparse data structures still incur significant overhead. We demonstrated that increasing MPI ranks per GPU mitigates serial bottlenecks, but rank scalability is restricted by GPU memory capacity and communication buffer overhead. To address these challenges, we proposed software optimizations targeting serial phases, such as reducing string-based metadata lookups, improving buffer cache management, and restructuring Kokkos kernels to reduce memory footprint. These insights pave the way for more efficient deployment of Parthenon and similar frameworks on upcoming exascale heterogeneous supercomputers, ultimately enabling faster and more accurate scientific simulations.

\section*{Acknowledgments}
We would like to thank Anthony Carreon and Jagmohan Singh for insightful discussions on Adaptive Mesh Refinement and the AMReX framework.
This material is based upon work supported by the Department of Energy, National Nuclear Security Administration under contract number 89233218CNA000001. LA-UR-25-29032 Approved for public release; distribution is unlimited.
\clearpage

\bibliographystyle{IEEEtranS}
\bibliography{reference}

\end{document}